\documentstyle[aps,pre,multicol,graphicx]{revtex}

\def\openbar{\rule{0mm}{2.5mm}
{\hfill\rule[-2.8mm]{0.2mm}{3mm}\rule{0.49\textwidth}{0.2mm}}
\vspace*{-5.5mm}
}

\begin{document}
\date{\today}
\title{A multibaker map for shear flow and viscous heating}
\author{
L\'{a}szl\'{o} M\'{a}ty\'{a}s,$^{(1)}$
Tam\'{a}s T\'{e}l,$^{(1)}$, and 
J\"{u}rgen Vollmer$^{(2,3)}$
}
\address{
(1) Institute for Theoretical Physics, E\"{o}tv\"{o}s University, P.
O. Box 32, H-1518 Budapest, Hungary.
\\
(2) Fachbereich Physik, Univ.-GH Essen, 45117 Essen, Germany.
\\
(3) Max Planck Institute for Polymer Research,
   Ackermannweg 10,
   55128 Mainz, Germany.
}
\maketitle

\begin{abstract}
  A consistent description of shear flow and the accompanied viscous
  heating as well the associated entropy balance is given in the
  framework of a deterministic dynamical system. A laminar shear flow
  is modeled by a Hamiltonian multibaker map which drives velocity and
  temperature fields. In an appropriate macroscopic limit one recovers
  the Navier-Stokes and heat conduction equations along with the
  associated entropy balance. This indicates that results of
  nonequilibrium thermodynamics can be described by means of an
  abstract, sufficiently chaotic and mixing dynamics. A thermostating
  algorithm can also be incorporated into this framework.
\end{abstract}

\draft

\begin{multicols}{2}
\section{Introduction}

Shear flows provide one of the paradigms of transport processes
\cite{GM,B97,books,Dorf,Focus}. The importance of chaos in the
equations underlying macroscopic shearing has recently been addressed
by various numerical studies \cite{ECM93,CherLeb97,shear,Wagner00},
which to some extent were supported by kinetic theory
\cite{books,Wagner00} and rigorous mathematical work
\cite{CherLeb97,Bunimovich}. In contrast, however, a simple, exactly
solvable model based on a low-dimensional chaotic dynamics --- whose
mixing property would be the cause of irreversibility --- has not yet
been established.  For material and heat transport such type of models
have helped to understand the physical content of thermostating
schemes used in numerical simulations \cite{MorRon96,Gasp,TV00,thesis}
(cf.~however \cite{CR2000} for open questions). In the present article
we introduce a similar model for shear flows in the hope that it can
also serve such a purpose.  The approach will be based on multibaker
maps. Previous works {in this spirit} successfully described the
phenomena of diffusion \cite{G,TG,Gent}, conduction in an external
field \cite{TVB,VTB97,VTB98,GD}, chemical reactions \cite{GasKla98},
thermal conduction \cite{TG98}, and cross effects due to the
simultaneous presence of an external field and heat conduction
\cite{MTV00,VTM00}.

\begin{figure}
\[ \includegraphics[width=0.3\textwidth]{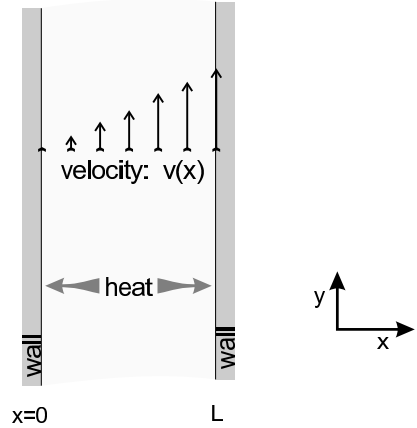} \]
\caption[]{\narrowtext
  Graphical illustration of {the} shear flow. The system
  is confined between two walls at positions $x=0$ and $L$.  The walls
  move relative to each other in the $y$ direction, thus inducing a
  velocity profile $v(x)$ indicated by vertical arrows. For a
  fixed temperature at the walls this leads to an accompanying heat
  flow in the $x$ direction (gray arrows).
\label{fig:geometry}}
\end{figure}

Our aim is to model a sheared fluid confined between two parallel
walls at the coordinates $x=0$ and $x=L$ (Fig.~\ref{fig:geometry}).
The flow is assumed to be two-dimensional in the $(x,y)$ plane. Shear
is induced by prescribing different $y$ components of the average
velocities ${\bf v}$ of particles close to the respective walls. In
order to make the calculations more transparent, we confine the
discussion to cases where the driving is sufficiently weak to induce
only a laminar flow, i.e., to cases where the velocity of particles is
always directed in the vertical direction such that ${\bf
  v}=(0,v(x))$.  For this system we establish a local entropy balance
that covers time dependent effects and does not rely on the
implementation of boundary conditions.

Three different boundary conditions for dealing with the
dissipated heat are considered:
{\bf (i) } In the simplest case the system is isolated.
A stationary linear velocity profile emerges in that case, and the
temperature becomes uniform. No steady state is reached, however, due
to a constant increase of temperature in response to the viscous
heating.
In addition, we consider systems where {\bf (ii) } there is a bulk
thermostat uniformly taking out the viscous heat, and {\bf (iii) } the
temperature is fixed {to the same value at both} boundaries 
so that the asymptotic
temperature profile is stationary, but no longer uniform.

The paper is organized as follows:
In Sect.~\ref{sec:thermodynamics} we recall basic notions of
irreversible thermodynamics that are to be { recovered} in a suitable
continuum limit of the multibaker dynamics.
In Sect.~\ref{sec:multibaker} we introduce the model, and
establish the evolution equations for the velocity and the
temperature field.
This allows us to address the entropy dynamics and its balance
equation (Sect.~\ref{sec:entropy}).
Subsequently, in Sect.~\ref{sec:limit} the macroscopic limit of
the resulting equations is worked out. 
The global behavior at different boundary conditions is compared to
irreversible thermodynamics in Sect.~\ref{sec:MacFlow}, and
conclusions are drawn in Sect.~\ref{sec:conclude}.

\section{Irreversible thermodynamics}
\label{sec:thermodynamics}

In this section we recall the thermodynamic description of
shear flows accompanied by viscous heating. The picture is
simplified by considering an incompressible fluid at constant
pressure.

\subsection{Transport equations}

For a system with constant density and pressure the temperature $T$ is the 
only relevant state variable, and for a complete description one also has 
to specify the velocity field ${\bf v}$ of the fluid \cite{GM,B97}. The 
thermodynamic state variables are the velocity field ${\bf v}$ and the 
temperature $T$ \cite{B97}. Mass conservation is expressed by a continuity 
equation.  For incompressible fluids the uniform mass density $\rho$ 
implies that the flow is divergence free, i.e., 
\begin{equation}
   \partial_i v_i = 0 .
\end{equation}
Here $i=x,y$ labels the components of the local flow velocity
   ${\bf v} \equiv (v_x, v_y)$,
and we adopted the Einstein convention, i.e., summation over
repeated indices.
The equation of motion for the velocity components is given by the
Navier-Stokes equation.
For the case of a negligible pressure gradient it reads
\begin{equation}
   \frac{d }{d t} v_i = \nu \partial_j \partial_j v_i ,
\end{equation}
where $\nu$ is the kinematic viscosity,
and $d/dt$ is the total time derivative.

The system of equations is closed \cite{B97} by the equation 
\begin{equation}
   \frac{d }{d t} T
 =
   \frac{\lambda}{\rho c_V} \partial_i \partial_i T
 + \frac{1}{2} \frac{\nu}{ c_V}
   (\partial_{k} v_l + \partial_{l} v_k)
   (\partial_{k} v_l + \partial_{l} v_k) 
\label{eq:ptTlong}
\end{equation}
for the temperature evolution. Here, $c_V$ is the
specific heat at constant volume, and $\lambda$ is the thermal
conductivity.

\subsection{Entropy balance}

The balance equation of the entropy density $s$ is
\begin{eqnarray}
   \partial_t s &=& \sigma^{(irr)} + \Phi  ,
\label{eq:s-dot}  \\
   \Phi &=& - \partial_i j^{(s)}_i + \Phi^{(th)}  ,
\label{eq:jsphith}
\end{eqnarray}
where $\sigma^{(irr)}$ is the irreversible entropy production
reflecting the viscous heating of the flow, $\Phi$ the entropy
flux, and ${\bf j}^{(s)}$ is the entropy-current density. The term
$\Phi^{(th)}$ models an entropy flux let into a heat
bath. In the bulk of a hydrodynamic system this term typically
vanishes, but it takes non-vanishing values wherever there is a
flux into the environment (i.e., for instance at the boundaries when
there is a heat flux through the walls).
In simulations where one does not desire to focus on the effects of
temperature gradients, and applies a thermostating algorithm, a
non-vanishing $\Phi^{(th)}$ can appear even in the bulk of the system.
We say that a system is subjected to an {\em ideal thermostat\/}, when
the dissipated heat is directly let into the heat bath.  In that case
the entropy current vanishes in the steady state, $j_i^{(s)}=0$, while
$\Phi^{(th)}$ is non-zero in the bulk, and counterbalances the
steady-state entropy production.

For an incompressible fluid in local equilibrium the Gibbs
relation
\(
   {\rm d} s = {{\rm d} u}/{T} 
\)
applies locally.
The evolution equations can be used to evaluate the terms in
Eq.~(\ref{eq:s-dot}). A straightforward calculation yields for the rate of
irreversible entropy production
(cf.~for instance \cite[Ch.~XII.~(23)]{GM})
\begin{equation}
   \sigma^{(irr)}
 =
   \lambda \frac{(\partial_i T)(\partial_i T)}{T^2}
+
   \frac{\nu \rho}{2T} 
   (\partial_{j} v_k + \partial_{k} v_j)
   (\partial_{j} v_k + \partial_{k} v_j) .
\label{eq:sirrlong}
\end{equation}
The associated entropy current takes the form \cite[Ch.~XII.~(22),
(24)]{GM}
\begin{equation} \label{eq:js}
   j^{(s)}_i = - \lambda \frac{\partial_i T}{T} .
\end{equation}
It depends only on the local temperature and its gradient. The
flow velocity ${\bf v}$ does not enter.

\subsection{Laminar Flow}

For a laminar flow driven by prescribed non-trivial $y$ components of
the velocity at the two walls, the velocity field at any position
$(x,y)$ takes the form ${\bf v}\equiv(0,v(x))$ (see
Fig.~\ref{fig:geometry}).  The $x$ component of the velocity vanishes,
and the profile is translational invariant in the $y$ direction
(parallel to the walls).  We restrict our investigation to cases where
the same holds for the temperature such that $T=T(x)$.  Consequently,
for the laminar flow the transport equations take the form
\begin{mathletters}
\begin{eqnarray}
   \partial_t v
&=&
   \nu \, \partial_x^2 v ,
\label{eq:ptv'}
\\
   \partial_t T
& =&
   \frac{\lambda}{\rho c_V} \, \partial_x^2 T
 + \frac{\nu}{c_V} \, (\partial_x v)^2 ,
\label{eq:ptTshort}
\end{eqnarray}
\end{mathletters}%
while the rate of irreversible entropy production $\sigma^{(irr)}$ and
the entropy current $j^{(s)}$ can be written as
\begin{mathletters}
\begin{eqnarray}
   \sigma^{(irr)}
& = &
   \lambda {\left( \frac{\partial_x T}{T} \right)}^2
 + \frac{\nu \rho}{T}  {(\partial_x v)}^2  ,
\label{eq:sirrshort}
\\
   j^{(s)}
&=&
   - \lambda \; \frac{\partial_x T}{T} . 
\label{eq:jsshort}
\end{eqnarray}
\end{mathletters}%
The scalar current $j^{(s)}$ denotes the $x$ component of the entropy
current, 
and an analogous convention is adopted for all currents.
The $y$ components of the currents vanish in the considered
setting.

\begin{figure}
\[ \includegraphics[width=0.45\textwidth]{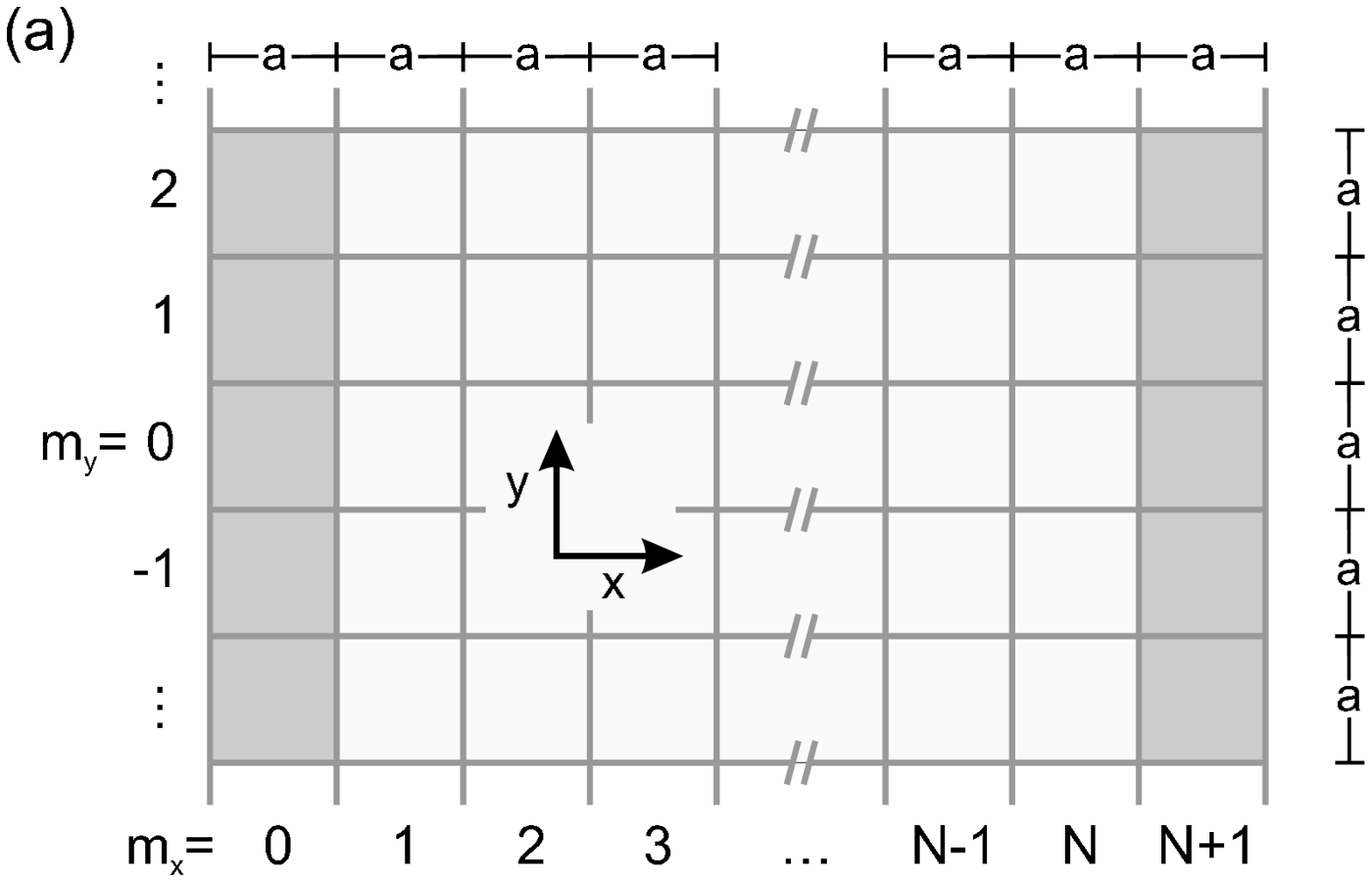} \]
\[ \includegraphics[width=0.45\textwidth]{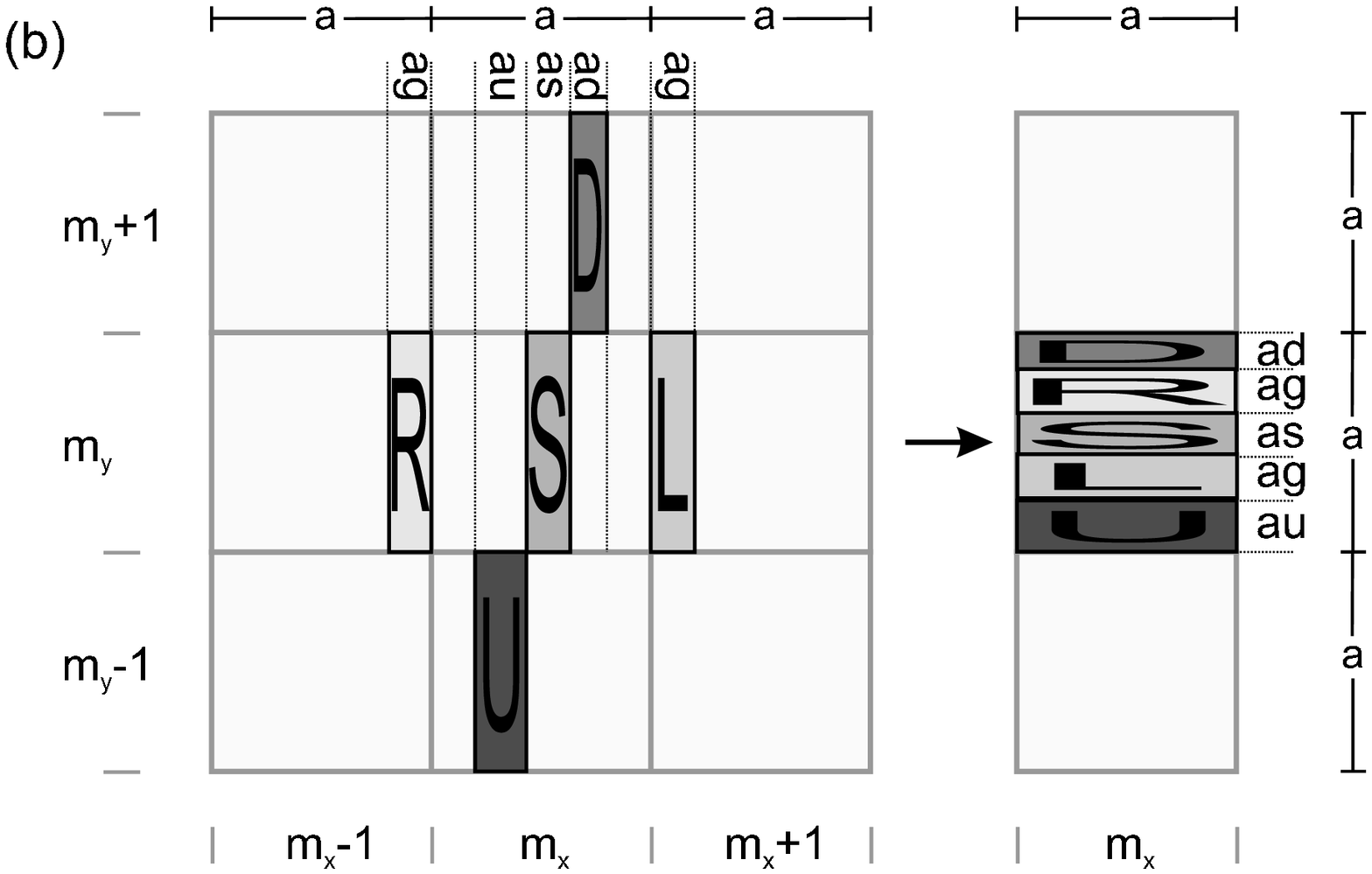} \]
\caption[]{\narrowtext
Graphical illustration of the action of the multibaker map on its
phase space  $(x,y)$. The indices labeling the columns and rows
are given on the lower and left margin, respectively, while their
width is indicated at the upper and right margins.
(a) The mapping is defined on a domain of $N+2$ identical columns of
square cells of size $a \times a$ that are labeled by the indices
$m_x=0,\cdots,N+1$ and $m_y=-\infty,\cdots,\infty$. Boundary
conditions on the flow are implemented in the shaded cells $m_x=0$ and
$N+1$.
(b)
The action of the map on regions that are mapped into cell
$(m_x,m_y)$. The contraction and expansion for these regions is
indicated by the deformation of the tags (R,U,S,D,L) in the
different branches of the map.
\label{fig:mbaker2D}}
\end{figure}
\section{The multibaker map}
\label{sec:multibaker}

In order to model the laminar flow we divide the plane into square
cells of size $a\times a$ that are large enough, on the one hand, to
admit thermodynamically meaningful averages, but, on the other hand,
sufficiently small to neglect gradients across cells.  The cells are
considered as the regions used in irreversible thermodynamics to
define local equilibrium variables.  The system shown in
Fig.~\ref{fig:geometry} is represented by a rectangular array of $N
\equiv L/a$ cells in horizontal direction, and an infinite number of
cells vertically (Fig.~\ref{fig:mbaker2D}a).  Cells are labeled by the
indices $m_x=0,\cdots,N+1$, and $m_y=-\infty,\cdots,\infty$.  All cells
have the same dynamics except the outermost ones where it is modified
to implement boundary conditions.

\subsection{Action of the mapping in the $(x,y)$ plane}

After each time unit $\tau$ every cell is divided into five columns
(Fig.~\ref{fig:mbaker2D}b). The rightmost (R) and leftmost (L) column have
width $ag$. They are mapped onto a strip of height $ag$ in cell
$(m_x+1,m_y)$ and cell $(m_x-1,m_y)$, respectively. The left column
(U) of widths $au$ is mapped upwards onto a strip of height $au$ in
cell $(m_x,m_y+1)$, and the right one labeled by (D) downwards into
$(m_x,m_y-1)$, respectively.  Region $S$ stays within the cell. In
all cases the area of the strips is preserved.

This dynamics is driving two fields. Denoting the composite index
$(m_x, m_y)$ as ${\bf m}$, these fields are
\\
(i) the velocity field $v_{\bf m}$ describing the mean flow velocity
$v(x,y)$ in cell ${\bf m}$ 
\\
(ii) an energy field $w_{\bf m}$ that represents the kinetic energy of
the fluid in cell ${\bf m}$.

The dynamics on this two-dimensional lattice can be considered as
a model of the velocity and energy transport in the configuration
space. In order to obtain a faithful representation of the entropy
balance, however, one has to consider the phase-space dynamics.
To define the dynamics in the analog of a $\mu$-space, we
take into account the translation invariance of the problem.
The velocity and energy fields can take on different values in the
columns $m_x = 0,\cdots,N+1$ of Fig.~\ref{fig:mbaker2D}a, but the
fields have to be uniform within every column. In this respect it is
not necessary to follow the dynamics in the $y$ direction, and one
achieves a quasi one-dimensional dynamics in an $(x,p)$ space, where
$p$ represents the non-trivial momentum-like variable of the model.

\subsection{Action of the mapping in the $(x,p)$ space}

\begin{figure}
\[ \includegraphics[width=0.45\textwidth]{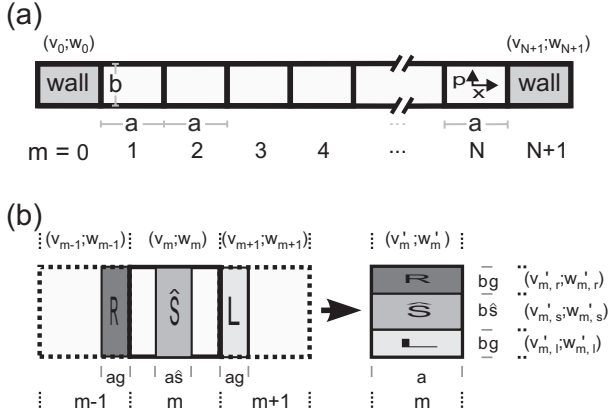} \]
\caption[]{\narrowtext
  Graphical illustration of the $(x,p)$ multibaker representation
  of the laminar flow.
  (a) Domain of the map. The outermost cells (indicated as grey boxes)
  are again used to implement boundary conditions.
  (b) Action in the bulk. The average values of the fields $v(x,y)$
  and $w(x,y)$ on the cells [cf. Eqs.~(\ref{eq:v-average}) and
  (\ref{eq:w-average})] and strips [cf. Eqs.~(\ref{eq:v_update}) and
  (\ref{eq:wlsrprime})] are given on the margins.
\label{fig:mbaker}}
\end{figure}

The domain of the multibaker in the $(x,p)$ space is shown in
Fig.~\ref{fig:mbaker}a. It comprises a chain of $N+2$ cells of
size $a\times b$, that for sake of more condensed notations are
labeled by the index $m$. The middle $N$ cells represent the bulk,
and two additional ones are used to implement boundary conditions
(cf.~Section \ref{sec:BC}). The parameter $b$ sets the momentum scale.
It will not play any role in thermodynamic considerations.

In order to maintain the same dynamics in the transport direction
in the two representations of the flow, each cell is divided into three 
columns of size $ag$, $a\hat{s}$ and $ag$, where $g$ is the same as above 
and $\hat{s}=1-2g$. The left and right columns of cell $m$ are mapped into 
a strip of height $ag$ in cell $m+1$ and cell $m-1$, respectively, as 
shown in Fig.~\ref{fig:mbaker}b. The middle column of size $a\hat{s}$ is 
squeezed and stretched onto a strip of height $a\hat{s}$ and remains in 
the same cell. 

This multibaker dynamics drives the velocity and energy fields.  Their
values might depend on the phase-space coordinate. Hence, we are
dealing with the bivariate {distribution of the velocity $v(x,p)$
  and kinetic energy $w(x,p)$ within each cell $m$. Only the cell
  averages, $v_m$ and $w_m$ appear in the transport equations. The
  dependence of $v(x,p)$ and $w(x,p)$ on the phase-space coordinates
  contributes essentially to the entropy dynamics, however.}

\subsection{The dynamics of the velocity field}

The mass density $\rho_m$ is proportional to the average phase-space 
density $\varrho_m$, in cell $m$, i.e., 
\begin{equation}
    \rho_m =  M b \varrho_m ,
\label{eq:rhomM}
\end{equation}
where $M$ is a constant of dimension mass. When interpreting the mass
density or associated quantities from the point of view of the
two-dimensional flow, $\rho_m$ is to be understood as a mass density
in the $x$ coordinate per unit length in $y$ direction.  Hence,
$\rho_m v_m$ represents the preserved momentum density (per unit
length) of the hydrodynamic flow.  Its evolution expresses momentum
conservation. As a consequence, the updated values for the velocity on
the strips $(L,\hat{S},R)$ after one time unit are (cf.~the right side
of Fig.~\ref{fig:mbaker}b)
\begin{eqnarray}
   \rho_m' v_{m,l}' & = & \rho_{m+1} v_{m+1} ,
\nonumber\\
   \rho_m' v_{m,s}' & = & \rho_m v_{m} ,
\label{eq:v_update}
\\
   \rho_m' v_{m,r}' & = & \rho_{m-1} v_{m-1} ,
\nonumber
\end{eqnarray}
where, $v_{m,l}'$, $v_{m,s}'$, and $v_{m,r}'$ stand for the
velocities of the flow in the regions $(L,\hat{S},R)$ of cell $m$,
respectively. The prime indicates that the updated values of the
field are considered.
The average momentum $\rho_m v_m$ of the full cell is the average
of the contributions on the different strips, i.e., 
\begin{equation}
\rho_m v_m
      =  (g \rho_{m,r} v_{m,r} + \hat{s} \rho_{m,s} v_{m,s} + g \rho_{m,l} v_{m,l})
\label{eq:v-average}
\end{equation}
at any instant of time. If the velocity is initially uniform in
the full cell (i.e., the values of $v$ in all strips agree), then
due to Eqs.~(\ref{eq:v-average}) and (\ref{eq:v_update}) the average velocity of
cell $m$ becomes after one time step
\begin{equation}
\rho_m' v_m'
      = (1-2g) \rho_m v_m + g \rho_{m-1} v_{m-1}  + g \rho_{m+1} v_{m+1} .
\end{equation}
Observing that the mass density $\rho$ is uniform and constant in
time, the updated value {of the} cell velocity $v_m$ is found to
be
\begin{eqnarray}
   v_{m}'
 &=&
   v_{m} +g (v_{m-1}+v_{m+1}-2v_m ) .
\label{eq:vn}
\end{eqnarray}
This evolution can be written in the form of a discrete balance
equation
\begin{equation}
   \frac{v_m' - v_m}{\tau}
 = - \frac{j_{m+1}^{(v)} - j_m^{(v)}}{a}
\end{equation}
with the discrete current
\begin{equation}
   j_m^{(v)} = - \frac{a^2 g}{\tau} \, \frac{v_m - v_{m-1}}{a} .
\label{eq:jmv}
\end{equation}
In the boundary columns $m=0$ and $N+1$ the dynamics will be
augmented by force terms (Sect.~\ref{sec:BC}) in order to
fix the velocity to the constant values $v_0\equiv v_L$ and
$v_{N+1}\equiv v_R$, respectively, irrespective of the momentum
flowing into these cells.

\subsection{The dynamics of the energy field}
\label{ssec:Efield}

The energy of cell $m$ is obtained by integrating $e_m\varrho_m$
over the volume $ab$ of the cell. At any instant of time it is
the spatial average of the energies $e_{m,l}$, $e_{m,s}$ and
$e_{m,r}$ of the different strips,
\begin{equation}
   e_m = g e_{m,l} + \hat{s} e_{m,s} + g e_{m,r} .
\label{eq:e-average}
\end{equation}
{The difference between the energy $e$ and the translational
  specific kinetic energy $M v^2/2$ of the flow defines the specific
  kinetic-energy density $w \equiv e - M v^2/2$, whose macroscopic
  limit will be proportional to the local temperature.}  Therefore, on
the strip {$i=l,s,r$} in cell $m$ one observes kinetic-energy
densities $w_{m,i}$ that fulfill
\[
   e_{m,i} = \frac{M}{2} v^2_{m,i} + w_{m,i} ,
\]
while the {coarse-grained} kinetic energy $e_m$ obeys at the same time
\[
   e_m = \frac{M}{2} v^2_{m} + w_m .
\]
Using these definitions and the averaging rule (\ref{eq:e-average}) for
the energy one finds
\end{multicols}
\widetext
\begin{equation}
   w_m
=
   g w_{m,l} + \hat{s} w_{m,s} + g w_{m,r}
+
   \frac{M}{2} \left[
      g v_{m,l}^2 + \hat{s} v_{m,s}^2 + g v_{m,r}^2
      - \left( g v_{m,l} + \hat{s} v_{m,s} + g v_{m,r} \right)^2 
      \right] .
\label{eq:w-average}
\end{equation}
\openbar
\begin{multicols}{2}
\noindent
This shows that the average kinetic energy $w_m$ is not the
straightforward spatial average of the values $w_{m,i}$ on the strips. 
Rather intra-cell variations of the velocity field (i.e., a non-trivial 
distribution of the $v_{m,i}$) also contribute to $w_m$. 

For a thermally closed system the specific kinetic energy $w$ is
advected by the flow. The values on the
strips after one time unit are then 
\begin{eqnarray}
   w'_{m,l} & = & w_{m+1} ,
\nonumber\\
   w'_{m,s} & = & w_{m} ,
\label{eq:wlsrprime}
\\
   w'_{m,r} & = & w_{m-1} .
\nonumber
\end{eqnarray}
{From} this and the update (\ref{eq:v_update}) of the velocities the
updated kinetic energy is found to be
\begin{eqnarray}
w'_m &=& w_m + g (w_{m-1}+w_{m+1}-2 w_m) +
\nonumber \\
&+&
  \frac{g  M}{2}  \left[ {(v_{m-1} - v_m)}^2+{(v_{m+1} - v_m)}^2 \right]
\nonumber \\
&-&
  \frac{M}{2}  {\left[ a^2 g \frac{(v_{m-1}+v_{m+1}-2v_m)}{a^2} \right]}^2 .
\label{eq:Tpn0}
\end{eqnarray}
The terms proportional to $M$ describe the effect of viscous
heating in this discrete setting.

In order to model the action of a thermostat, that leads to local changes 
of the specific kinetic energy due to a heat flux into the environment, an 
additional {\em thermostat heat source \/} $q_m$ is incorporated into the 
update of $w_m$ by multiplying the right-hand side of Eq.~(\ref{eq:Tpn0}) by a 
factor $[1+\tau q_m]$, 
\begin{eqnarray}
w'_m
&=&
\biggl\{ w_m + g (w_{m-1}+w_{m+1}-2 w_m) +
\nonumber \\
&+&
  \frac{g  M}{2}  \left[ {(v_{m-1} - v_m)}^2+{(v_{m+1} - v_m)}^2 \right]
\nonumber \\
&-&
  \frac{M}{2} g^2 {\left[ v_{m-1}+v_{m+1}-2v_m \right]}^2
\biggr\} \; [1+ \tau q_m]  .
\label{eq:Tpn}
\end{eqnarray}
This equation can be rewritten as
\begin{equation}  
   \frac{w_m' - w_m}{\tau}
 = Q_m
 + \frac{a^2g}{\tau} \frac{w_{m+1} - 2 w_m + w_{m-1}}{a^2}
\label{eq:wmBalance0}
\end{equation}
with the {\em full heat source\/}
\begin{eqnarray}
   Q_m
 &=&
\frac{q_m}{1+\tau q_m} w_m'
\nonumber \\
 &+&
\frac{a^2 g}{\tau} \frac{M}{2}
\left[
\frac{{(v_{m-1} - v_m)}^2}{a^2}+\frac{{(v_{m+1} - v_m)}^2}{a^2}
\right]
\nonumber \\
&-&
\frac{\tau}{2} M
{\left[ \frac{a^2 g}{\tau} \frac{(v_{m-1}+v_{m+1}-2v_m)}{a^2}
\right]}^2 . 
\label{eq:Qmmicr}
\end{eqnarray}
The equation for the update of $w$ can be rearranged into a
balance equation for the heat per unit volume
\begin{equation}
   \rho \frac{w_m' - w_m}{\tau}
 = \rho Q_m
 - \frac{j_{m+1}^{(w)} - j_m^{(w)}}{a}
\label{eq:wmBalance}
\end{equation}
which comprises the divergence of the discrete ``heat'' current
\begin{equation}
   j_m^{(w)} = - \frac{a^2 g}{\tau} \;
      \frac{\rho}{M} \, \frac{w_m - w_{m-1}}{a} . 
\label{eq:jmw}
\end{equation}
The first contribution to $Q_m$ in (\ref{eq:Qmmicr}) reflects the action
of the thermostat, and the latter two the effect of viscous heating of
the fluid. A steady state with a uniform $w$ profile can be found for
$Q = 0$, which thus mimics an ideal thermostat. In contrast, in a bulk
system in the sense of conventional irreversible thermodynamics
$q_m=0$, and $Q_m$ only vanishes when the discrete velocity
gradients $v_{m+1}-v_m$ and $v_m-v_{m-1}$ across the cell vanish.

\section{Entropies and their time evolution}
\label{sec:entropy}

\subsection{The coarse-grained and the Gibbs entropy}

The Gibbs entropy for cell $m$ of a multibaker system with a
density field $\varrho(x,p$) and specific kinetic-energy field
$w(x,p)$ takes the form \cite{MTV00}
\begin{equation}
   S_m^{(G)}
 = - k_B
\int_{\mbox{\small cell }m} dx\,dp \; \varrho(x,p)
\ln\left[ \frac{\varrho(x,p)}{\varrho^*} w(x,p)^{-\gamma} \right]
\label{eq:defSG}
\end{equation}
where $\gamma$ is a free exponent.
Since the phase-space density $\varrho(x,p)$ is constant, one can
identify the reference density $\varrho^{*}$ with $\varrho(x,p)$, and
obtains
\begin{equation}
S_m^{(G)}  =  k_B \gamma  \frac{a\rho}{M}  
      \int_{\mbox{\small cell }m}  \frac{dx}{a} \,\frac{dp}{b}  \; \ln w(x,p) .
\label{eq:SmG}
\end{equation}
Here, we have already replaced the phase-space density by the mass
density (\ref{eq:rhomM}). 

The coarse-grained entropy of cell $m$ is based on the
cell-averaged coarse-grained value $w_m$ of the kinetic energy
$w(x,p)$. For the form  (\ref{eq:SmG}) of the Gibbs entropy, one
obtains
\begin{equation}
   S_m= k_B a  \gamma \frac{\rho}{M} \ln w_m . 
\end{equation}
as the coarse-grained entropy.

In order to discuss 
the thermodynamic time evolution of entropies, one conveniently 
starts with uniform densities in every cell \cite{VTB98,thesis}.
In that case the coarse-grained and the Gibbs entropy initially
coincide. After one time step, however, they typically differ.  The
Gibbs entropy has changed due to the fact that the $w$ field takes
different values on the strips $(R,\hat{S},L)$, leading to the new
value
\end{multicols}
\widetext
\begin{eqnarray}
   S_m^{(G)}
 &=& k_B \gamma \varrho a b \;
     \left\{
         g \; \ln[ w_{m-1} \; (1+\tau q_m) ]
       + (1-2g) \; \ln[ w_m \; (1+\tau q_m) ]
       + g \; \ln[ w_{m-1} \; (1+\tau q_m) ]
     \right\}
\nonumber \\
 & = &
    k_B \gamma \varrho a b
    \left\{ \; \ln[ w_m \; (1+\tau q_m) ]
  +  \, g \; \ln\frac{w_{m-1}}{w_m}
  -  \, g \; \ln\frac{w_m}{w_{m+1}} \right\} .
\end{eqnarray}
\openbar
\begin{multicols}{2}
\noindent
On the other hand, the coarse-grained entropy after one time step is
\begin{equation}
   S'_m = - k_B a b \gamma \varrho \ln w_m' .
\end{equation}

\subsection{Entropy balance}

The coarse-grained entropy only depends on the average kinetic-energy
density $w_m$, and is therefore considered as the multibaker analogue
of the thermodynamic entropy.  Its temporal change can be decomposed
as
\begin{eqnarray}
  \frac{\Delta S_m}{a b \tau}
& \equiv &
   \frac{S_m' - S_m}{a b \tau}
\nonumber\\
& = &
   \frac{( {S_m}'- {S_m^{(G)}}' )  -  ( S_m - S_m^{(G)} )}
        {a b \tau}
+
   \frac{{S_m^{(G)}}' - S_m^{(G)}}{a b \tau} ,
\label{eq:micbal}
\end{eqnarray}
where information-theoretic arguments \cite{VTB97,VTB98,MTV00,thesis}
lead one to identify
\begin{mathletters}
\begin{equation}
  \frac{\Delta_i S_m}{a b \tau}
\equiv
   \frac{( {S_m}'- {S_m^{(G)}}' )  -  ( S_m - S_m^{(G)} )}
        {a b \tau}
\label{eq:DeltaSiDef}
\end{equation}
with the rate of entropy production, and
\begin{equation}
  \frac{\Delta_e S_m}{a b \tau}
\equiv
   \frac{{S_m^{(G)}}' - S_m^{(G)}}{a b \tau}
\label{eq:DeltaSeDef}
\end{equation}
\label{eq:DeltaSDef}%
\end{mathletters}%
with the entropy flux. Note that the second term of the numerator of
$\Delta_i S_m$ vanishes due to the initial condition of uniform
fields in the cells.

Inserting the values for $S_m'$ and $S_m^{(G)'}$ into the
definition (\ref{eq:DeltaSiDef}) yields
\begin{eqnarray} \label{eq:DiSm}
   \frac{\Delta_i S_m}{a \tau}
&=&
   \frac{k_B \gamma \rho}{M \tau}
   \biggl[
      \ln\left( \frac{w_m'}{w_m} (1+\tau q_m)^{-1} \right)
\\ 
&-&
   g \ln \frac{w_{m-1}}{w_m} - g \ln \frac{w_{m+1}}{w_m} \biggr] .
\nonumber 
\end{eqnarray}
Remarkably, this expression for the entropy production does not
depend on the source term $Q_m$, but only on the values of the
coarse-grained field $w$ in cell $m$ and its neighbors.  
{The shear flow enters only indirectly through the update $w_m'$ of
  $w_m$.}

Similarly to the other balance equations, the entropy flux 
can be written as
\begin{equation}  \label{eq:DeSm}
   \frac{\Delta_e S_m}{a \tau}
=
   - \frac{j_{m+1}^{(s)} - j_{m}^{(s)}}{a}
   + \Phi_m^{(th)} ,
\end{equation}
where the discrete entropy current $j_m^{(s)}$ takes the form
\begin{mathletters}
\begin{equation}
   j_m^{(s)}
 = -  k_B \frac{a \, g}{\tau} \; \gamma \frac{\rho}{M} \;
   \ln\frac{w_{m+1}}{w_m}  ,
\label{eq:jmsDef}
\end{equation}
and
\begin{equation}
   \Phi^{(th)}_m = k_B \gamma \frac{\rho}{M} q_m
\label{eq:PhiTh}
\end{equation}
\end{mathletters}%
describes the direct flux into the thermostat.
Such a flux is encountered whenever there is a non-vanishing source
term $q_m$.  In view of Eq.~(\ref{eq:Qmmicr}) this finding further supports
the interpretation of $q_m$ and $Q_m$ given at the end of
Sect.~\ref{ssec:Efield}.

\section{The macroscopic limit}
\label{sec:limit}

In this section we evaluate the expressions of the different
quantities considered in Sects.~\ref{sec:multibaker} and
\ref{sec:entropy}, and identify conditions for consistency with the
thermodynamic results described in Sect.~\ref{sec:thermodynamics}.

\subsection{Definition of the limit}
\label{sec:def_limit}

The macroscopic limit implies that $L\gg a$ (i.e., $N\gg 1$), and
$\tau$ is much smaller than typical macroscopic time scales. Formally
it can be defined as
\begin{equation}
   a,\tau \rightarrow 0
\end{equation}
such that the spatial coordinate
\begin{eqnarray}
   x &\equiv& a m
\end{eqnarray}
is finite. As mentioned earlier, the field $w$ is assumed to go over into 
the local temperature $T(x)$ in the macroscopic limit, i.e., 
\begin{eqnarray}
   w_m & \rightarrow & C k_B T(x) ,
\label{eq:CkbT}
\end{eqnarray}
where $C$ is a dimensionless constant.

\subsection{The transport equations}

The macroscopic form of the
velocity current (\ref{eq:jmv}) becomes
\begin{equation}
    j_m^{(v)} = - \frac{a^2 g}{\tau} \partial_x v .
\label{eq:jmv-limit}
\end{equation}
In order to achieve a meaningful thermodynamic result
the ratio $a^2 g/\tau$ has to be finite in the limit.
Indeed, comparison with Eq.~(\ref{eq:ptv'}) shows that
\begin{equation}
  \nu \equiv \frac{a^2 g}{\tau}
\label{eq:ga2}
\end{equation}
is the kinematic viscosity.

Equations (\ref{eq:jmv-limit}) and (\ref{eq:ga2}) imply that Eq.~(\ref{eq:vn}) is
the discrete form of the Navier-Stokes equation (\ref{eq:ptv'})
for the laminar flow.  Moreover, with this choice for $g$ one
obtains for the heat current (\ref{eq:jmw})
\begin{equation}
j^{(w)} = - \frac{\rho}{M} \; \nu C k_B \partial_x T .
\end{equation}
Hence, Fick's law of heat conduction is recovered with the heat
conduction coefficient
\begin{equation}  \label{eq:lambda1}
   \lambda =   \frac{\rho}{M} \; \nu C k_B .
\end{equation}

Due to Eq.~(\ref{eq:lambda1}), the macroscopic limit of Eq.~(\ref{eq:Tpn})
reduces to Eq.~(\ref{eq:ptTshort}) for the temperature evolution:
\begin{equation}
   \partial_t T
=
   \frac{\lambda}{C k_B \rho/M} \partial_x^2 T
 + \frac{\nu M}{C k_B}  {(\partial_x v)}^2 + q T . 
\label{eq:ptTmacr}
\end{equation}
Comparing the coefficients in this equation with the ones in
Eq.~(\ref{eq:ptTshort}), one obtains
\begin{equation} 
C k_B = c_V M . 
\end{equation} 
The proportionality constant $C$ introduced in Eq.~(\ref{eq:CkbT}) corresponds thus 
to the specific heat at constant volume (measured in units of $k_B$).

\subsection{The entropy balance}

In the macroscopic limit the rate of irreversible entropy production
(\ref{eq:DiSm}) becomes
\begin{equation}
   \frac{\Delta_i S_m}{a \tau}
 \rightarrow
   \sigma^{(irr)}
 =
   k_B \gamma \nu \frac{\rho}{M} \left( \frac{\partial_x T}{T} \right)^2
 + \frac{\nu \rho}{T}  {(\partial_x v)}^2  .
\label{eq:DeltaSmIrr}
\end{equation}
It fully agrees with the thermodynamic form of the entropy production 
(\ref{eq:sirrshort}) when the coefficient $k_B \gamma \nu \rho/M$ in 
front of the first term is the heat conductivity. In view of 
(\ref{eq:lambda1}) we thus conclude, that in our model $\gamma = C$, i.e., 
the exponent $\gamma$ appearing in the definition (\ref{eq:defSG}) of the 
entropy is proportional to the specific heat. Hence, the final form of the 
heat conductivity $\lambda$ can be settled {to}
\begin{equation}
   \lambda = \frac{\rho \gamma k_B \nu}{M} = \rho c_V \nu .
\label{eq:lambda2}
\end{equation}

Working out the expression (\ref{eq:jmsDef}) of the entropy current
$j_m^{(s)}$, one finds in the macroscopic limit
\begin{equation}
   j^{(s)} (x)
= - k_B \gamma \nu \frac{\rho}{M} \frac{\partial_x T}{T}
= - \lambda \frac{\partial_x T}{T} .
\end{equation}
In view of (\ref{eq:lambda2}) this relation also fully agrees
with its thermodynamic counterpart (\ref{eq:jsshort}).

\section{Boundary Conditions}
\label{sec:BC}

In this section we demonstrate how boundary conditions for the
shear flow can be implemented in the multibaker dynamics. 

\begin{figure}
\[ \includegraphics[width=0.48\textwidth]{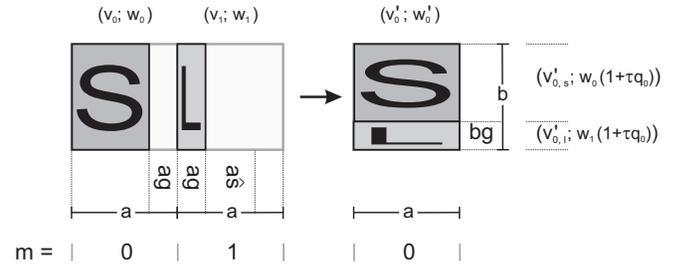} \]
\caption[]{\narrowtext
Implementation of the boundary condition in cell $0$. The
indication of the width of strips, values of the fields on the
strips, and action of the map follows the rules spelled out in
Fig.~\ref{fig:mbaker}.
\label{fig:bc}}
\end{figure}

\subsection{Boundary conditions for the velocity field}

The {multibaker} dynamics for cell $0$ is shown in Fig.~\ref{fig:bc}. (An
analogous prescription holds in cell $N+1$.) The action of the strip
mapped from cell $0$ to $1$ fully agrees with the one of the
corresponding strips in the bulk such that the dynamics of cell $1$
{agrees with} the bulk dynamics of Fig.~\ref{fig:mbaker}b. The
important difference about the dynamics at the boundary is that
\\
(i) there is no column leaving cell $0$ to the left, and
\\
(ii) the momentum of the particles entering cell $0$ is not preserved.
\\
Condition (i) reflects that no particles can penetrate the walls of
the system, and (ii) reflects that there is a force exerted by the
boundary on the fluid. In strip $L$ entering cell $0$ (and analogously
strip $R$ entering cell $N+1$) the update of the velocity consequently
contains the contribution of the external force $f_0$ ($f_{N+1}$),
leading to
\begin{eqnarray}
   v'_{0,l} & = & v_{1} + \frac{f_0 \tau}{g \rho} , 
\nonumber \\
   v'_{N+1,r} & = & v_{N} + \frac{f_{N+1} \tau}{g \rho} . 
\label{eq:wallveloc}
\end{eqnarray}
The update of the 
{coarse-grained} velocity of the leftmost cell is 
\begin{equation}
v'_0 = v_0 + g (v_1 -v_0) + \frac{f_0 \tau}{\varrho} ,
\end{equation}
and {an analogous relation holds for cell $N+1$.}  
A shear flow with a prescribed shear rate is enforced by adjusting
the forces in such a way that $v_{0}$ and $v_{N+1}$ take the
respective values $v_L$ and $v_R$ of the velocities at the walls at
any time. 

In view of the time evolution of $v_m$,  
{in a steady state [Eq.~(\ref{eq:vn}) with $v'_m=v_m$], 
the asymptotic 
velocity profile becomes linear
\begin{equation}
  v_m 
= v_L + \frac{m}{N+1} \; (v_R - v_L) 
\equiv v_L + \frac{m}{N+1} \; \Delta v 
\label{eq:vm-distr}
\end{equation}
irrespective} of the temperature distribution in the system.

\subsection{Boundary conditions for the kinetic-energy density}
\label{ssec:BCw}

The modification of the dynamics of the velocities at the boundary
also implies changes of the thermostat source terms.
Since the velocity in cell $0$ is uniform by definition, the update of $w_0$
can be found from Eq.~(\ref{eq:w-average}) where the terms containing
$w_{m,l}$ and $v_{m,l}$ are not present since no particles enter
cell $0$ from the left. Consequently,
\[
   w_0'
=
   \left[
      w_0 + g ( w_{1}-w_0 )
   \right]
   ( 1 + \tau q_0 ) .
\]
This leads to the discrete temporal change
\begin{equation}
   \frac{w_0' - w_0}{\tau}
=
   \frac{q_0}{1 + \tau q_0} w_0'
 + \frac{1}{a} \; \frac{a^2 g}{\tau} \; \frac{ w_{1}-w_0 }{a}
\label{eq:w0update}
\end{equation}
instead of Eq.~(\ref{eq:wmBalance0}). For a steady state ${w_0' - w_0}$ vanishes, and
the equation implies [cf.~Eq.~(\ref{eq:jmw})]
\begin{equation}
   a q_0
=
   - \frac{a^2 g}{\tau} \;
     \frac{(w_1 - w_0)/a }{ w_0 + g (w_1-w_0) }
=
   - \frac{ j_1^{(w)} }{\varrho \; [w_0 + g (w_1-w_0)] } . 
\label{eq:q0}
\end{equation}
The right-hand side of this expression is finite in the
macroscopic limit, where 
\begin{equation}
   a q_0 
\rightarrow  - \nu  \left. \frac{\partial_x T}{T} \right|_{x=0}
       =     - \frac{\nu \; j^{(T)}(0)}{\lambda \; T(0)}  .
\label{eq:q0limit}
\end{equation}
Since, however, $a \rightarrow 0$ in the macroscopic limit, the
thermostat heat source $q_0$ cannot be interpreted as a density when
there is a finite heat current $j^{(w)}(0)$. Its integral over the
cell, $a q_0$, stays however finite. As a consequence, the heat
current has to vanish in a steady state when $q_0 = q_{N+1} = 0$.

\section{Different macroscopic flows}
\label{sec:MacFlow}

We have seen that in the macroscopic limit the local momentum, energy
and entropy balances for the multibaker map coincide with their
thermodynamic forms. This result was achieved by inspecting the local
time evolution of the densities without referring to particular
boundary conditions, and it is not restricted to steady states. In
order to underline these features, we discuss the global transport
pictures for three different settings of transport.  We emphasize,
that the considerations immediately generalize to arbitrary
time-dependent states although the particular calculations are carried
out for stationary coarse-grained velocity fields.

\subsection{Isolated systems}

In thermodynamics sheared systems are often considered to
be thermally isolated. In that case $q_m$ identically vanishes for all
cells $m=0,\cdots,N+1$. 
Eq.~(\ref{eq:w0update}) implies for the current (\ref{eq:jmw})
\begin{equation}
j^{(w)}_1 \equiv - \frac{a^2 g}{\tau} \frac{\rho}{M} \frac{w_1-w_0}{a} 
          = - \frac{a \rho}{M} \frac{w'_0-w_0}{\tau}  . 
\end{equation}
Assuming that the time derivative of $w_0$ is finite in the 
macroscopic limit, 
{the current 
$j^{(w)}_1 \rightarrow - a (\rho/M) C k_B \partial_t T(0)$ approaches }
zero as $a\rightarrow 0$. 
An analogous arguments applies also to $j_{N+1}^{(w)}$. 

Thus, for the {asymptotic state,} the boundary conditions imply
that $w_0$ and $w_{N+1}$ practically coincide with $w_1$ and
$w_N$, respectively.  Moreover, the viscous heating term $Q_m$ in the
bulk only depends on the square of $v_{m+1}-v_m$.  The heating is
therefore spatially uniform in the long time limit.  With this input,
one immediately verifies that $w_m$ approaches a spatially uniform
value $w^*$.  However, in view of the heat source $Q_m$ in
(\ref{eq:Qmmicr}), it constantly grows in time,
\begin{equation} 
  \frac{w^{*'} - w^*}{\tau}
= \nu M \; \left( \frac{\Delta v}{a (N+1)} \right)^2 .
\end{equation} 
This {\em temporal\/} evolution of $w$ reflects the rise of
temperature due to the entropy production in the system. After all,
\[
   \frac{\Delta_i S_m}{a \tau}
 \equiv
   \frac{k_B \gamma \rho}{\tau M} \; \ln\frac{w^{*'}}{w^*}
 =
   \frac{k_B \gamma \rho}{w^*} \; \nu \;
\left( \frac{\Delta v}{a (N+1)} \right)^2 
 + {\cal O}(\tau)
\]
which in the macroscopic limit reduces to the thermodynamically
expected value
\[
  \sigma^{(irr)}(t) =  \frac{\rho \, \nu}{T^*(t)} \;
{\left( \frac{\Delta v}{L} \right)}^2 
\]
where $T^*(t) = \hbox{const} + (\nu/c_V) { (\Delta v/L) }^2 t $ is the
spatially uniform, but temporally increasing asymptotic temperature
distribution. Note that the flux $\Phi^{(th)}$ and the entropy current
vanishes in this setting, implying an ever increasing entropy.

\subsection{Systems subjected to an ideal (bulk) thermostat}

In a system subjected to an ideal thermostat the viscous heating
is instantaneously released into a heat bath. Accordingly, $Q_m$
vanishes for all cells $m=1,\cdots,N$ such that $q_m$ can be
determined from Eq.~(\ref{eq:Qmmicr}). 

Moreover, according to the results obtained in Sect.~\ref{ssec:BCw} we
are again dealing with a system with vanishing heat currents through
its boundaries, i.e., for the asymptotic state the boundary conditions
imply $w_0=w_1$ and $w_N=w_{N+1}$, and there is no source term $Q$ for
the heat. Consequently, $w_m$ approaches a spatially uniform state
$w^*$ that is {\em stationary\/} in this case. Entropy production
arises due to the nontrivial form of $q_m$ [cf.~Eq.~(\ref{eq:Qmmicr})
with $Q_m=0$]
\[
   q_m \equiv q^* = - \frac{\nu M \; \left(\frac{\Delta v}{a(N+1)}\right)^2}
                {w^* + \nu M \tau \; \left(\frac{\Delta v}{a(N+1)}\right)^2}
\]
in the bulk [cf.~the steady-state velocity profile
(\ref{eq:vm-distr})]. Hence, the irreversible entropy change is
\begin{eqnarray}
   \frac{\Delta_i S_m}{a \tau}
 &\equiv&
   -\frac{k_B \gamma b\varrho}{\tau} \; \ln(1 + \tau q^*)
\nonumber\\
 &=&
   \frac{k_B \gamma \rho}{M w^*} \; \nu \; \frac{\Delta v^2}{a^2 (N+1)^2}
 + {\cal O}(\tau)  ,
\end{eqnarray}
and in the macroscopic limit it reduces to the thermodynamically
expected value
\[
\sigma_m^{(irr)} =
   \frac{\rho \, \nu}{T^*} \; {\left( \frac{\Delta v}{L} \right)}^2 ,
\]
where $T^*$ is the steady-state temperature. Also in this case there
is no entropy current in the steady state. However, at every location
in the system there is a non-vanishing entropy flux into the
thermostat
\begin{equation}
\label{eq:Phithisol}
\Phi^{(th)} = k_B \gamma b\varrho q_m
            \rightarrow  - \frac{\rho\nu}{T^*}
            {\left( \frac{\Delta v}{L} \right) }^2 ,
\end{equation}
which exactly compensates $\sigma^{(irr)}$ such that the entropy
indeed is 
stationarity. 

\subsection{Thermostating at the walls}

Finally, we discuss a steady state in a hydrodynamic bulk system that
generates heat flux into the walls due to the prescribed temperatures
at the boundaries. There are no thermostat heat sources in the bulk,
i.e., $q_m=0$ for $m=1,\cdots,N$. On the other hand, there are source
terms $q_0$ {and} $q_{N+1}=q_0$ acting in the two outermost cells
{that} fix the values of $w_0$ and $w_{N+1}$ (i.e., the temperature)
to the same constant value $w_0$. To this end the sources
counterbalance a macroscopic heat current releasing the viscous heat
into the bath at the boundaries [cf.~Eq.~(\ref{eq:q0})]. In contrast to the
previous cases, the discrete heat current does not vanish at the
boundaries. In view of Eqs.~(\ref{eq:jmw}) and (\ref{eq:w0update}) the heat current
through the left boundary of the steady state system is
\begin{equation}
\frac{j_1^{(w)} }{\rho} 
   \equiv - \nu \; \frac{w_1 - w_0}{a}
      = \frac{a q_0}{1 + \tau q_0} \; w_0 .
\end{equation}
Since the dynamics is symmetrical and $w_{N+1}=w_0$, the
current at the right boundary takes the same value up to a change
of the sign
\[
   j_{N+1}^{(w)} =  - j_{1}^{(w)} .
\]
In a steady state the cells in the bulk consequently fulfill 
\[
   0 =  \rho \frac{w'_m - w_m}{\tau}
     = \rho \nu \; \frac{\Delta v^2}{a^2 (N+1)^2}
       - \frac{j_{m+1}^{(w)} - j_{m}^{(w)}}{a} ,
\]
{leading} to a linear profile of the heat current,
\begin{equation}
\label{eq:jmwlin} 
   j_{m}^{(w)}
 =
   a \left( m - 1 - \frac{N}{2} \right) \;
   \rho \nu \; \left( \frac{\Delta v}{a(N+1)} \right)^2 .
\end{equation}
In view of the definition Eq.~(\ref{eq:jmw}) of $j_m^{(w)}$, this implies a
parabolic profile of $w$ in the steady state, 
\begin{equation}
\label{eq:wmparab}
   w^*_m = w_0 -
   \frac{am}{2} \; a [m - (N+1)] \; M \;
   \left( \frac{\Delta v}{a(N+1)} \right)^2 ,
\end{equation}
as expected for the temperature profile in a
shear flow subjected to identical temperatures at the two ends.

The entropy production in the bulk is related to the spatial
variation of 
the steady-state distribution $w^*_m$ [cf.~Eq.~(\ref{eq:DiSm})],
\begin{eqnarray}
   \sigma_m^{(irr)}
 &\equiv&
   -\frac{k_B \gamma \rho}{\tau M} \;
   \left[ g \ln\frac{w^*_{m-1}}{w^*_m} + g \ln \frac{w^*_{m+1}}{w^*_m} \right]
\nonumber\\
 &=&
   \frac{k_B \gamma}{M} 
   \left[
      \frac{1}{w^*_m} \; \frac{j_{m+1}^{(w)} - j_{m}^{(w)}}{a}
      + \frac{{j_{m+1}^{(w)}}^2 + {j_{m}^{(w)}}^2}{2 \rho \nu w_m^{*2}}
   \right]   ,
\end{eqnarray}
where the second equation was obtained by expanding the logarithms to
second order in the differences $w^*_{m+1}-w^*_m$, and rearranging
terms.  Using Eqs.~(\ref{eq:wmBalance}), (\ref{eq:Qmmicr}) and
(\ref{eq:jmwlin}), the first term can be related to the applied shear
rate, while the second one represents the contribution to the entropy
production arising from the heat flow.  Indeed, in the macroscopic
limit one recovers in this case both contributions to
Eq.~(\ref{eq:DeltaSmIrr}).  (For small shear rates the contribution
from the temperature change is, however, negligible).
Correspondingly, there is a finite entropy current $j^{(s)}$ at every
point in the system, but only at the boundaries there is a flux into
the thermostat.  According to Eq.~(\ref{eq:q0limit}) the macroscopic
limit of the full entropy flux through the boundaries becomes
\begin{equation}
a\Phi^{(th)} = k_B \gamma b\varrho a q_0 
\rightarrow k_B \gamma \nu \frac{\rho}{M}  
\left. \frac{\partial_x T}{T} \right|_{x=0} .  
\end{equation}
Since the stationary temperature profile is obtained from (\ref{eq:wmparab}) in 
the form of 
   $T(x) = T(0) - [x \, (x-L) \, (M /2 C k_B)]  (\Delta v /L )^2 $, 
the derivative at the origin is 
   $\left. \partial_x T \right|_{x=0}= M L (\Delta v /L )^2 /(2Ck_B) $. 
The contribution at the right boundary is the 
same. Therefore, we find for the integrated entropy flux 
\begin{equation} 
   \Phi^{(th,tot)} 
= 
   - \frac{\rho\nu}{T_0} L {\left( \frac{\Delta v}{L} \right)}^2 ,  
\end{equation}
where we used $\gamma=C$. For weak shear this is essentially the same
as the integral of the constant flux (\ref{eq:Phithisol}) over the
chain. Thus, we conclude that for sufficiently small heat currents
thermostating in the bulk and in the boundaries can lead to the {\em
  same\/} global behavior.

The work per unit time done by the external force densities $f_0$ and
$f_{N+1}$ can be evaluated using their form Eq.~(\ref{eq:wallveloc})
and the value of the constant velocity gradient
\begin{equation}
   a(f_0 v_L +f_{N+1} v_R) = \nu \rho \frac{(\Delta v)^2}{L}  .  
\end{equation}
This expression is exactly $- T_0 \Phi^{(th,tot)}$, such that 
{in a steady state the work done by the external forces equals the
  total heat flux into the thermostat.}

\section{Discussion}
\label{sec:conclude}
We have enlarged the family of multibaker maps by a model for
momentum, energy and entropy transport in viscous hydrodynamic flows.
Although the macroscopic problem is of strongly dissipative nature,
the proposed multibaker dynamics is area preserving (Hamiltonian).
This is to be contrasted with previous multibaker models of electric
transport \cite{VTB97,VTB98} and thermoelectric cross effects
\cite{MTV00,VTM00} where the inclusion of a reversible dissipation
mechanism was necessary in order to simulate the effect of
thermostating on the particle dynamics. 
The form in which 
thermostating appears in the present model is via a heat-source term
in the microscopic energy dynamics (which was already present in the
model of cross effects \cite{MTV00}). This term, however, does not
give rise to phase-space contraction.

The model has the following basic features:
\\
(a) The time evolution of the system can be interpreted as that of
weakly interacting particles. The resulting ``multi-baker'' gas obeys
the classical ideal-gas equation of state. The particle and heat
diffusion coefficient, and the kinematic viscosity are proportional to
each other, as in the kinetic theory for classical ideal gases
\cite{Reichl}.
\\
(b) The distribution of a macroscopic velocity does not enter the
entropy explicitly. Rather the shear rate appears in the entropy
balance via its influence on the temperature dynamics only.
\\
(c) The connection to a thermodynamic description of transport is
achieved by considering fields, which are coarse grained in regions of
small spatial extension. Their properties are to be contrasted to
those of fields characterizing the microscopic evolution.
\\
(d) Comparing these two levels of description allows us to identify
all contributions to the {\em local\/} entropy balance, in full
consistency with thermodynamics.  They apply to both stationary and
transient states.
\\
(e) A source term is implemented in the evolution equation of the
microscopic kinetic energy.  {It} provides the possibility to
implement local irreversible cooling of the system, i.e., to extract
heat such that states which are permanently warmed up by viscous
heating can become stationary. In traditional thermodynamics these
terms are only present at the boundaries, and they vanish in the bulk.
\\
(f) It is indicated how the velocity and the kinetic-energy dynamics
can be implemented at the walls in order to achieve correspondence
with different macroscopic boundary conditions.
\\
(g) When source terms are present in the bulk, the local entropy
balance of nonequilibrium thermodynamics is generalized by
introducing at every location an instantaneous flux of entropy (i.e.,
of heat) into a thermostat (the entropy flux is then no longer the
divergence of the entropy current).  The dynamics is in that case
reminiscent of numerical algorithms based on Gaussian thermostats.
\\
(h) The global entropy balance was worked out, in order to demonstrate
that the total heat flux into a thermostat is independent to a large
extend of whether thermostating is applied in the bulk or at the
walls.

The major interest of the present model lies in the light it sheds on
the origin of viscous heating in deterministic models of transport. It
was pointed out how fractal structures emerge in multibaker models for
a variety of physical settings of shear flow.  In all cases the
structures arise from the mixing of regions with different local
temperatures and flow velocities whose differences are exponentially
proliferating to smaller and smaller scales for a nonequilibrium
system.  One can explicitly follow this redistribution of the kinetic
energy, until it reaches the scale of the domains used to define local
thermodynamic averages. Motion at that scale has to be considered as
contributing to the non-directional motion, hence leading to viscous
heating. It is {\em only\/} due to this coarse-graining mechanism that
the macroscopic shear rate appears in the expression of the
irreversible entropy production.

\section*{Acknowledgments}

We would like to thank Bob Dorfman, Christan Gruber, and Gary Morris for
illuminating discussions.
Support from the Hungarian Science Foundation (OTKA Grand No.~032423) is 
acknowledged. 

 

\end{multicols}

\end{document}